\documentclass[aps,prl,reprint,superscriptaddress]{revtex4-1}
\usepackage{amsmath}
\usepackage{color}
\usepackage[dvipsnames]{xcolor}
\usepackage[final]{changes}
\usepackage{setspace}
\definechangesauthor[color=Red]{MC}
\definechangesauthor[color=Blue]{FP}
\definechangesauthor[color=olive]{GD}
\definechangesauthor[color=Violet]{GLL}

\usepackage{todonotes}
\usepackage[normalem]{ulem}

\newcommand{\popn}{\langle c_n^\dagger c_n\rangle}
\newcommand{\popl}{\langle c_l^\dagger c_l\rangle}

\newcommand{\polvn}{\langle v_n^\dagger c_n\rangle}

\newcommand{\polcl}{\langle c_l^\dagger v_l\rangle}

\newcommand{\bs}{\langle b_s\rangle}
\newcommand{\Bs}{\langle b_s^\dagger\rangle}
\begin{document}
%\raggedbottom

%\preprint{}

\title{Reply to the Comment on ``Thermal, quantum antibunching and lasing
thresholds from single emitters to macroscopic devices"}
\author{Mark Anthony Carroll}
\affiliation{Department of Physics, University
of Strathclyde,  107 Rottenrow,  Glasgow G4 0NG, UK.}
\author{Giampaolo D'Alessandro}
\affiliation{School of \replaced[id=GD]{Mathematical Sciences}{Mathematics}, University of Southampton, Southampton SO17
1BJ, United Kingdom}
\author{Gian Luca Lippi}
\affiliation{Universit\'e C\^ote d'Azur, Institut de Physique de Nice, UMR 7710
CNRS, 1361 Route des Lucioles, 06560 Valbonne, France}
\author{Gian-Luca Oppo}\affiliation{Department of Physics, University
of Strathclyde,  107 Rottenrow,  Glasgow G4 0NG, UK.}
\author{Francesco Papoff}
%\email{f.papoff@strath.ac.uk}
\affiliation{Department of Physics, University
of Strathclyde,  107 Rottenrow,  Glasgow G4 0NG, UK.}

\date{\today}

\begin{abstract}
We deconstruct and address a comment to Carroll et al. [Phys Rev Lett 126, 063902 (2021)] (PRL) that has been posted on arXiv appearing as two versions [arXiv:2106.15242v1] and [arXiv:2106.15242v2].  This comment claimed that a term in the model presented in the PRL had been incorrectly omitted and that, hence, the laser threshold predicted by the model  in the PRL is unattainable.  We show that the term in question was correctly neglected because it represents collective effects that are not observable in the devices modelled in the PRL. Moreover, even if this term were to be included, the laser threshold would still be present, contrary to what was claimed in the comment. We conclude that the model presented in PRL is correct and that its results are innovative and of wide application in laser physics and quantum optics.
\end{abstract}

\maketitle

The Comment \cite{vyshnevyy2021commentv1,vyshnevyy2021commentv2} by A. A. Vyshnevyy and D. Yu. Fedyanin states that some terms in Eq. (2) in Ref. \cite{carroll2021thermal} have been omitted without justification.  Comment \cite{vyshnevyy2021commentv1,vyshnevyy2021commentv2} claims that the laser threshold emerging from a revised Coherent-Incoherent Model (CIM)~\cite{carroll2021thermal} is ``unattainable'' when the term $\sum_{n\neq l}\delta\langle c_l^\dagger v_l v_n^\dagger c_n\rangle$ is added to the equation for the photon assisted polarization $\delta \langle b c^\dagger v \rangle$. The authors claim that neglecting this term is not justified because \added[id=GD]{it} ``represents the macroscopic electric polarization of the gain medium in the same sense as $\delta \langle b^\dagger b \rangle$ represents the cavity field." \cite{vyshnevyy2021commentv1}.  Moreover, \replaced[id=GD]{the Comment}{it} identifies the classical polarization $|P|^2$ with $\sum_{n,l}\langle c_l^\dagger v_l v_n^\dagger c_n\rangle$, thus claiming that neglecting $\sum_{n\neq l}\delta\langle c_l^\dagger v_l v_n^\dagger c_n\rangle$ violates the quantum-classical correspondence. The Reply is straightforward: none of the statements and definitions of the relevant quantities given in \cite{vyshnevyy2021commentv1,vyshnevyy2021commentv2} are as claimed. In the following, we refer to the standard quantum optics interpretation (e.g. \cite{kira1999quantum}) of the quantities used in Ref.  \cite{carroll2021thermal}.

Here we show that: 1) The threshold exists, persists and is {\em attainable} even with the wrong assumptions of \cite{vyshnevyy2021commentv1,vyshnevyy2021commentv2}; 2) When correctly taking into account terms of the order of $\sum_{n\neq l}\delta\langle c_l^\dagger v_l v_n^\dagger c_n\rangle$ and the sum's spatial nonlocality, the CIM provides accurate values of the laser threshold.

We consider first the term $\sum_{n\neq l}\delta\langle c_l^\dagger v_l v_n^\dagger c_n\rangle$: since different emitters are in different positions, the sum does not correspond to the modulus squared of the macroscopic polarization {\bf $P(\boldsymbol{x})$}, a local function of position $\boldsymbol{x}$ in Maxwell's equations. Instead, it corresponds to the integral $\int_{|\boldsymbol{y-x}|>\epsilon} P(\boldsymbol{x})P^*(\boldsymbol{y}) d\boldsymbol{y}$, where $|\boldsymbol{y-x}|>\epsilon$ eliminates single emitter terms as does $n \ne l$ in the sum. \added[id=MC]{Furthermore, $|P|^2$ does not correspond to $\sum_{n,l}\langle c_l^\dagger v_l v_n^\dagger c_n\rangle$  as \added[id=GLL]{incorrectly} suggested in \cite{vyshnevyy2021commentv2}. Imposing operator normal ordering gives 
$\sum_{n,l}\langle c_l^\dagger v_l v_n^\dagger c_n\rangle =\langle c_l^\dagger c_l\rangle - \sum_{n,l} \langle c_l^\dagger v_n^\dagger v_l c_n\rangle \neq  |P|^2$
where $\langle c_l^\dagger c_l\rangle$ is the excited state population and $\sum_{n,l} \langle c_l^\dagger v_n^\dagger v_l c_n\rangle$ the sum of the expectation values of the product of polarisations between QDs placed at different positions: a spatially nonlocal term.  This decomposition proves the point. The term that corresponds to the amplitude squared of the local polarization from Maxwell's equations is instead the term $|\langle v^\dagger c \rangle|^2$, present in the CIM (see Eq. (2) in \cite{carroll2021thermal}) but arbitrarily and inconsistently removed from Eq. (1) in \cite{vyshnevyy2021commentv1,vyshnevyy2021commentv2}.}

In nanolasers, terms like $\sum_{n\neq l}\delta\langle c_l^\dagger v_l v_n^\dagger c_n\rangle$ are normally neglected. They represent collective effects \deleted[id=GD]{(CE)}, like superradiance, usually not observable in the presence of strong polarisation dephasing due to high carrier density screening \cite{gies2007semiconductor}. It is also shown in \cite{protsenko2021quantum} that in the limit of strong polarisation dephasing $(\gamma_c / \gamma) \ll1$ (where $\gamma_c$ and $\gamma$ are the cavity decay and dephasing rates, respectively) the correlations describing collective effects are negligible and that \deleted[id=GLL]{\replaced[id=GD]{collective effects}{CE}} \deleted[id=GLL]{are only important} \added[id=GLL]{they become important only} in low-Q cavities where ($\gamma_c/\gamma) \geq 1$.  The CIM~\cite{carroll2021thermal} matches the parameters of (GaAs-based) Quantum Dots most commonly used in nanolasers, with very rapid decays of the intrinsic polarization~\cite{sugawara2000effect} and negligible polarization correlations. \added[id=GD]{Therefore the elimination of this term in the CIM~\cite{carroll2021thermal} is perfectly justified.}

\added[id=GD]{On the contrary, the term incorrectly identified with $|P|^{2}$ in~\cite{vyshnevyy2021commentv1, vyshnevyy2021commentv2} is physically not realistic. To verify this we use the cluster expansion to write the dynamical equation for $\langle c_l^\dagger v_n^\dagger v_l c_n\rangle$. This is}
\begin{align}
  \label{eq:CVvc_EVfull}
     d_t \langle c_l^\dagger v_n^\dagger v_l c_n\rangle &=   -(2\gamma +
    i\Delta\varepsilon)\langle c_l^\dagger v_n^\dagger v_l c_n\rangle \\ \nonumber&+ g_{ls}^*\big[\langle b_s^\dagger v_n^\dagger
    c_n\rangle(1-2\langle c_l^\dagger c_l \rangle) \\ \nonumber
    & -2\polvn\langle b_s^\dagger c_l^\dagger c_l\rangle
    +2\Bs\popl\polvn \big]  \\ \nonumber &+ g_{ns}\big[\langle b_s c_l^\dagger v_l\rangle(1-2\langle
    c_n^\dagger c_n\rangle) \\ \nonumber & - 2\polcl\langle b_s c_n^\dagger c_n\rangle + 2\bs\popn\polcl\big],
\end{align}
where the coefficients $g_{ns}$ depend on the cavity-mode field at the QDs positions~\cite{baer2006}.  \added[id=MC]{ The phase difference between QDs is not an issue in terms with coupling coefficients and polarization operators that belong to the same QD, because of the phase invariance $g_{ls}\rightarrow g_{ls}e^{i\phi}, c_l^\dagger v_l\rightarrow c_l^\dagger v_le^{i\phi}$. However, as a result of
the spatial nonlocality, Eq.~(\ref{eq:CVvc_EVfull}) includes products of coupling coefficients, $g_{ls}$, and polarisation operators, $v_n^\dagger c_n$,
corresponding to different QDs. Neglecting these phase
differences, as done in \cite{vyshnevyy2021commentv1,vyshnevyy2021commentv2}, implies that all QDs are identical and the products of the field mode with the dipole
interband matrix elements of all QDs (see definition of
$g_{ls}$ in \cite{baer2006}), have identical phase and amplitude.} These extremely strict conditions cannot be satisfied by all QDs for physically realistic boundary conditions.

\added[id=GD]{Having dealt with the incorrect identification of $|P|^{2}$ we now focus on the comment's second major claim, namely that ``the predicted bifurcation point [is] unattainable''.  This is simply not correct.}  Adding Eq. (1) of~\cite{vyshnevyy2021commentv1,vyshnevyy2021commentv2} to CIM~\cite{carroll2021thermal} displaces the \replaced[id=GD]{bifurcation point, i.e. the lasing threshold}{threshold} (black star in Fig.1) from its original position~\cite{carroll2021thermal} (blue star), rendering the post-bifurcation dynamics unstable due to the arbitrary removal of terms of comparable size to $\sum_{n\neq l}\delta\langle c_l^\dagger v_l v_n^\dagger c_n\rangle$. Consistently computing (as in \cite{carroll2021thermal}) the variables at the appropriate order (cf. Eq.~(\ref{eq:CVvc_EVfull}) above), but keeping the unphysical assumption of identical QD coefficients~\cite{vyshnevyy2021commentv1, vyshnevyy2021commentv2} stabilizes the dynamics, moving the threshold to a lower pump value (red star). Finally, relaxing the unphysical condition of identical QD coefficients, the threshold returns to approximately the CIM value (red diamond and cross).  In summary: thresholds leading to coherent fields can always be observed\added[id=GLL]{!} Contrary to claims~\cite{vyshnevyy2021commentv1, vyshnevyy2021commentv2}, the model of \cite{carroll2021thermal} is correct and widely applicable.

\begin{figure}
  \centering
  \includegraphics[clip=false,width=.5\textwidth]{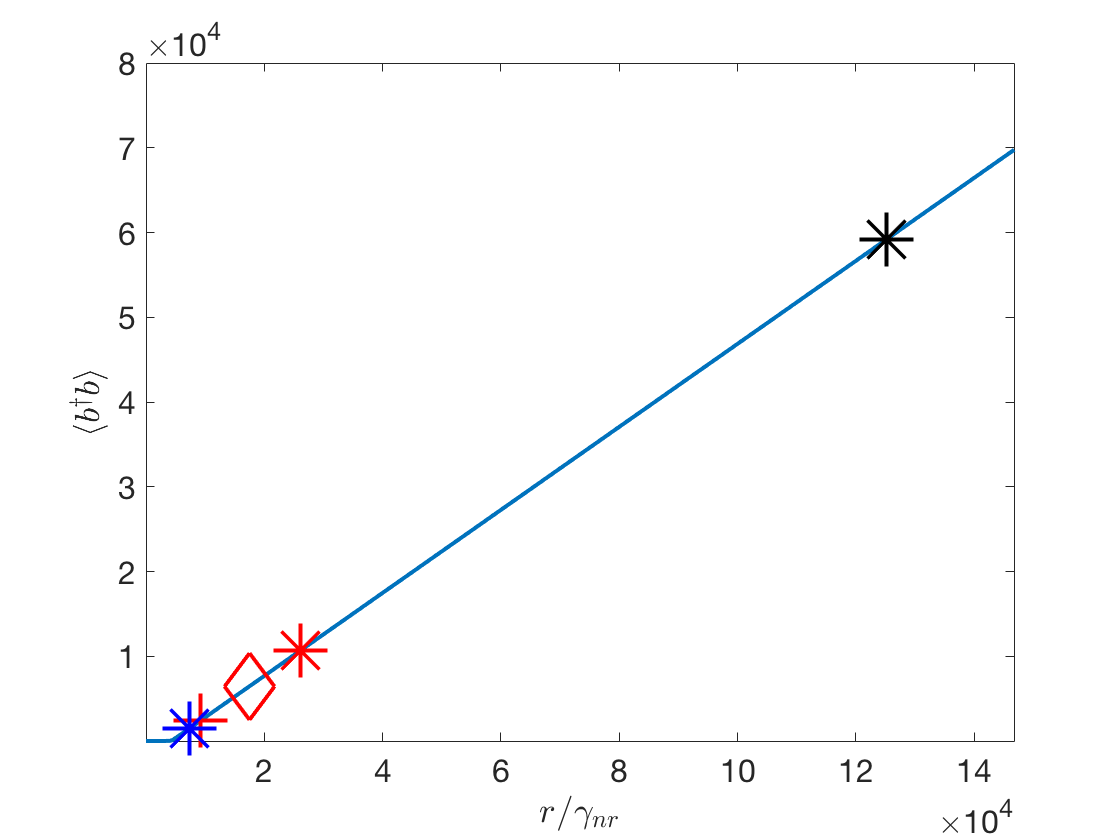}
  \caption{ \label{fig:comparison} 
Photon number versus pump for 40 QDs. The blue star is the laser threshold, $r\approx7.3\times10^{12}s^{-1}$, of the CIM \cite{carroll2021thermal};  the black star of the CIM plus Eq. (1) of \cite{vyshnevyy2021commentv1,vyshnevyy2021commentv2}, $r\approx1.3\times10^{14}s^{-1}$; the red star when variables ignored in \cite{vyshnevyy2021commentv1,vyshnevyy2021commentv2} are included, $r\approx2.6\times10^{13}s^{-1}$; the red diamond (red cross) assumes that only $90\%$ ($50\%$) of the QDs and of their coupling coefficients are identical, $r\approx1.7\times10^{13}s^{-1}$ ($r\approx9.2\times10^{12}s^{-1}$). All parameter values are the same as in~\cite{carroll2021thermal} and $\gamma_{nr}=10^9s^{-1}$. }
\end{figure}

%%%%%%%%%%%%%%%%%%%%%%%%%%%%%%

Having now established the validity of the CIM \cite{carroll2021thermal} and \replaced[id=GLL]{pointed out the errors in} {and the invalidity of} the model proposed in \cite{vyshnevyy2021commentv1,vyshnevyy2021commentv2}\added[id=GLL]{,} we correct and clarify the outstanding claims and statements made \replaced[id=GD]{in the Comment}{by the authors}. \added[id=MC]{The term $\delta \langle b^\dagger b \rangle$ is not the cavity field as stated in \cite{vyshnevyy2021commentv1}.
$b^\dagger b$ is the photon number operator and its expectation value $\langle
b^\dagger b \rangle$ corresponds to the  intensity of the resonant cavity mode,
{\bf not to the cavity field}. The term corresponding to the (complex) amplitude of
the coherent laser mode is $\langle b \rangle$.
From the cluster expansion $\langle b^\dagger b \rangle = \delta \langle
b^\dagger b \rangle + \langle b^\dagger \rangle \langle b \rangle$, one sees
that the term $ \delta \langle b^\dagger b \rangle $ is the difference
between the total photon number and the coherent fraction which possesses a
collective phase.} 

\added[id=GLL]{Contrary to a claim made in~\cite{vyshnevyy2021commentv1}, the}
\deleted[id=GLL]{The} model presented in \cite{carroll2021thermal} is not a ``Maxwell-Bloch'' model\deleted[id=GLL]{ as claimed in \cite{vyshnevyy2021commentv1}}. It is a quantum model derived directly from a quantum Hamiltonian and naturally describes spontaneous emission. On the contrary, \added[id=GLL]{the Comment~\cite{vyshnevyy2021commentv2} follows a semiclassical perspective when including, in an {\it ad hoc} manner, quantum fluctuations to mimic spontaneous emission.} \deleted[id=GLL]{even though Eq.~(5) in \cite{vyshnevyy2021commentv2} includes quantum fluctuations by \replaced[id=GD]{incorporating}{including} spontaneous emission, it does this in an ad-hoc manner, following a semiclassical perspective.} \replaced[id=GLL]{Thus, its approach~\cite{vyshnevyy2021commentv2}}{Therefore, the approach taken in the Comment \cite{vyshnevyy2021commentv2}} is not self consistent.

\added[id=GLL]{It is important to remark that neglecting}
\deleted[id=GLL]{Neglecting} $\delta \langle b^\dagger b c^\dagger c\rangle$ and $\delta \langle b^\dagger b v^\dagger v\rangle$ is standard procedure with cluster expansions~\cite{fricke96a} at the two-particle level \cite{gies2007semiconductor} \added[id=GLL]{and that Ref. \cite{carroll2021thermal} follows the traditional approach of previous publications on this issue}.

\added[id=GLL]{To conclude with miscellaneous mistakes in \cite{vyshnevyy2021commentv1,vyshnevyy2021commentv2} we remark}
\added[id=MC]{\deleted[id=GLL]{Note also} that Jaynes-Cummings is spelled incorrectly as ``Janes-Cummings"} and that Eq. (5) in \cite{vyshnevyy2021commentv1} is \added[id=GLL]{erroneously given and should instead} read: 
\begin{align}
\langle c_l^\dagger c_l\rangle_{th} = \frac{1}{2} +
\frac{\gamma_c\gamma}{2N|g|^2}\Bigg[1 + \Big(\frac{\Delta\nu}{\gamma_c +
\gamma}\Big)^2\Bigg]
\end{align}
where the nonlasing state is stable for $\langle c^\dagger c\rangle<\langle
c^\dagger c\rangle_{th}$ and unstable if $\langle c^\dagger c\rangle>\langle
c^\dagger c\rangle_{th}$. 

Finally, there is a \added[id=GD]{possible} misinterpretation regarding the emission after the bifurcation in~\cite{vyshnevyy2021commentv1, vyshnevyy2021commentv2}\replaced[id=GD]{. The characterisation of the bifurcation as ``abrupt transition to lasing at a finite
pump rate'' is misleading}{}: close to threshold only a fraction of the photon field is coherent and single-frequency.

%\bibliography{bib}


																		\end{document}